\documentclass[twocolumn,showpacs,preprintnumbers,superscriptaddress,amsmath,amssymb,prl]{revtex4}

\usepackage{graphicx}
\usepackage{dcolumn}
\usepackage{bm}
\usepackage{amsmath}
\usepackage{indentfirst}

\newcommand{\be}{\begin{equation}}
\newcommand{\ee}{\end{equation}}
\newcommand{\ba}{\begin{eqnarray}}
\newcommand{\ea}{\end{eqnarray}}

\begin{document}

\preprint{APS preprint}

\title{Andrade, Omori and Time-to-failure Laws from Thermal Noise in
Material Rupture}

\author{A. Saichev}
\affiliation{Mathematical Department,
Nizhny Novgorod State University, Gagarin prosp. 23,
Nizhny Novgorod, 603950, Russia}
\affiliation{Institute of Geophysics and Planetary Physics,
University of California, Los Angeles, CA 90095}

\author{D. Sornette}
\affiliation{Institute of Geophysics and Planetary Physics,
University of California, Los Angeles, CA 90095}
\affiliation{Department of Earth and Space Sciences, University of
California, Los Angeles, CA 90095\label{ess}}
\affiliation{Laboratoire de Physique de la Mati\`ere Condens\'ee,
CNRS UMR 6622 and Universit\'e de Nice-Sophia Antipolis, 06108
Nice Cedex 2, France}
\email{sornette@moho.ess.ucla.edu}

\date{\today}

\begin{abstract}
Using the simplest possible ingredients of a rupture model with thermal
fluctuations, we provide an analytical theory of three ubiquitous
empirical observations obtained in creep (constant applied stress)
experiments: the initial Andrade-like and Omori-like $1/t$ decay of the
rate of deformation and of fiber ruptures and the $1/(t_c-t)$ critical
time-to-failure behavior of acoustic emissions just prior to the
macroscopic rupture. The lifetime of the material is controlled by a
thermally activated Arrhenius nucleation process, describing the
cross-over between these two regimes. Our results give further credit to
the idea proposed by Ciliberto et al. that the tiny thermal fluctuations
may actually play an essential role in macroscopic deformation and
rupture processes at room temperature. We discover a new re-entrant effect
of the lifetime as a function of quenched disorder amplitude.
\end{abstract}


\maketitle

Constant stress (so-called ``creep'') experiments
constitute a standard testing procedure
in material sciences. The typical
response to the sudden application of a constant stress
is that the strain rate as well as the acoustic emission rate first jump
to high values followed by slow universal power law decays, respectively 
called
the Andrade law \cite{Andrade}
for the strain rate and the Omori law \cite{Omori} for the acoustic rate.
Then, after a long decay whose duration may vary within
extraordinary large bounds (see below), the rates rebound and accelerate
(while the applied stress remains constant)
by following a power law acceleration resulting in a finite-time singularity
(the rupture of the sample).
The two regimes of decelerating followed by accelerating rates and
the lifetime of the structure are the result of a subtle interplay
between the pre-existing micro-heterogeneity of the material and the
self-organized evolving deformation and damage due to dislocation motion 
and/or
micro-cracking. Up to now, there are no theory encompassing
all these regimes. Here, we propose a simple mechanism that provides
an explanation of all these observations, which is based on the recent
proposal \cite{cilietal1,Politi} that thermal noise is strongly
renormalized by quenched heterogeneities. Based on the analysis of a simple
fiber bundle rupture model, Refs.\cite{cilietal1,Politi} showed that the
average lifetime of the fiber-bundle takes an Arrhenius
form with an effective temperature renormalized from the
bare temperature $T$ to a value strongly amplified by the
presence of the frozen disorder in the rupture thresholds
$f_c(i)$, in agreement with experiments and numerical
simulations. This result suggests that the usual assumption
of neglecting the role of thermal fluctuations in material
rupture processes at room temperature may actually be
incorrect (see \cite{early} for early discussions): due to
frozen heterogeneities, tiny thermal fluctuations can be
amplified many times, thus actually controlling the
time-dependent aspects of failure. Our purpose is to
extend the analysis of this model by showing that it is
able to reproduce all the empirical observations mentioned
above in creep (constant applied stress) experiments. This
shows that the simplest possible ingredients of a rupture
model together with thermal fluctuations can render
essentially all of the richness of creep experiments.

The democratic fiber-bundle model (DFBM) with thermal noise
\cite{cilietal1,Politi} can be seen
as a mean field treatment of rupture. A macroscopic constant load
$F=N f_0$ is applied
at time $t=0$ to an initially undamaged system made of a very large number
$N$ of parallel elastic fibers (the results derived below
are obtained in the thermodynamic limit $N \to \infty$).
At all times, $F$ is shared democratically
among all $(1-\Phi(t)) N$ surviving fibers, where $\Phi(t)$ is the fraction
of broken fibers at time $t$. The externally applied force per
surviving fiber is thus
\be
f_a=\frac{f_0}{1-\Phi(t)}\,.
\label{fa}
\ee
The strength of each fiber $i$ is characterized by a critical
value $f_c(i)$ drawn for a distribution $P_d(f)$, centered on the
mean equal to $1$
and with variance $T_d$. Putting the mean strength to $1$ sets the force 
scale. The heterogeneous strengths are constant and determined once for all,
corresponding to a frozen disorder, which is ``read'' in a certain
organized way as the rupture develops.
Microscopic thermal fluctuations are taken into account by assuming that
a fiber with load $f_a$ and threshold $f_c(i) > f_a$ has a non-zero
probability $G(f_c(i)-f_a)$ to rupture per unit time governed by the
rate with which
a thermal fluctuation can activate a microscopic force $\Delta f_i
\geq f_c(i)-f_a$
to pass the rupture threshold $f_c(i)$:
$G(f_c(i)-f_a)= \frac{\gamma}{2}\, \text{erfc}\left(
\frac{f_c(i)-f_a}{\sqrt{2T}}\right)$, where $ \text{erfc}(x)$ is
the complementary error function, $T$ is the
variance of the thermal force fluctuations $\Delta f_i$ and $\gamma$ is a 
microscopic
constant rate fixing the time scale of the thermal activation process.
This expression amounts to introducing a zero-mean normal distribution
of thermal fluctuation forces $\Delta f_i$ with variance $T$ and with
correlation time proportional to $1/\gamma$.

We first follow \cite{Politi} and introduce
the distribution $Q(f,t)$ of the rupture thresholds of the unbroken
fibers at time $t$. Obviously, $Q(f,t)=0$
for $f<f_a$, since all these fibers
are already broken. Politi et al. \cite{Politi} have shown that $Q(f,t)$ can
be approximated with a very high accuracy in the limit $N\to\infty$ by
the initial distribution $P_d(f)$ of rupture strengths truncated at a lower
value $f_s(t)$,
\be
Q(f,t) = P_d(f)~,~~{\rm for}~f>f_s(t)
\label{fgjaa}
\ee
and $0$ otherwise, where $f_s(t)$ is determined by the self-consistent 
equation
\be
\Phi(t) = \int_{-\infty}^{f_s(t)} df~P_d(f)
\label{condself}
\ee
expressing that all fibers whose strengths are below $f_s(t)$ have failed
at some time before $t$. This approximation for $Q(f,t)$ with 
(\ref{condself})
amounts to view the time-dependent rupture as a ``front'' propagating
and ``eating'' the distribution $P_d(f)$ from the weakest towards the
strongest fibers.
We also have by definition
$\Phi(t) = 1 -\int_{-\infty}^{+\infty} df~Q(f,t)$.
Taking the time derivative of $\Phi(t)$
and replacing $\dot{Q}(f,t)$ by
$-Q(f,t)~G(f-f_a)$ expressing that
the rate of breaking is controlled
the thermally activated rupture process acting on each fiber independently,
we get $\dot{\Phi} = \int_{-\infty}^{+\infty} df~Q(f,t)~G(f-f_a)$.
Putting (\ref{fgjaa}) in this equation and taking for $P_d(f)$
a normal distribution centered on $1$
with variance $T_d$ as in \cite{cilietal1,Politi} yields
\be
\dot{\Phi}=\frac{\gamma}{2}\int_{f_s}^\infty
\frac{1}{\sqrt{2\pi T_d\mathstrut}}\exp\left[-
\frac{(1-f)^2}{2 T_d}\right]\,\text{erfc} \left(
\frac{f-f_a}{\sqrt{2T}}\right)df~.
\label{mflwsl}
\ee
Making explicit $P_d(f)$ in (\ref{condself}) gives
\be
\Phi=\frac{1}{2}\left[\text{erf} \left(\frac{f_s-1}{\sqrt{2
T_d\mathstrut}}\right)+1\right],~
f_s=1+\sqrt{2T_d}~\text{irf~}(2\Phi-1),
\label{gjkwl}
\ee
where $y=\text{irf\,}(z)$ is the inverse function to the
error function $z=\text{erf\,}(y)$.
Putting  $f_s$ in (\ref{mflwsl}) gives
\be
\dot{\Phi}=R(\phi) \equiv \frac{\gamma}{2}\int_\Phi^1
\text{erfc}\left[L(\Phi,z)\right]dz\,,
\label{mgler}
\ee
\be
L(\Phi,z)= \frac{1}{\sqrt{2T}}
\left(1-\frac{f_0}{1-\Phi}\right) +\mu\,
\text{irf\,}(2z-1)\,,
\label{mmld}
\ee
with $\mu=\sqrt{T_d/T}$. The solution of equation (\ref{mgler})
with (\ref{mmld}) provides in principle all the information on the
fraction $\Phi(t)$
of broken fibers.
This equation (\ref{mmld}) is valid as long as the approximation
(\ref{fgjaa}) holds (see below).

The parameter $\mu$ quantifies the relative importance of the thermal
fluctuations compared with the quenched heterogeneities. The
relevant regime for applications
to macroscopic ruptures at room temperature is $\mu > 1$ and often
$\mu \gg 1$, that is,
thermal fluctuations are tiny contributions to the applied
macroscopic mechanical forces.
Indeed, assuming that the energy barrier to rupture a fiber corresponds
to the Griffith energy $\approx g~c^2$ necessary for nucleating a
crack of half-length $c$
in the solid with surface energy $g$, we obtain
$\mu \approx 1.5-4 \cdot 10^3$ for $c=1$ micron and
$\mu \approx 1.5-4$ for $c=1$ nanometer, using
$g=10-50$ erg/cm$^2$ for most solids. Thus, even for the smallest 
microcracks,
thermal fluctuations are very small in relative value.

It turns out that this regime $\mu \gtrsim 1$ allows for a very convenient
approximation of $R(\Phi)$ obtained
by linearizing $L(\Phi,z)$ with respect to $z$. Then, the integral
over $z$ in (\ref{mgler}) can be calculated explicitly to yield
\be
\dot{\Phi}= R(\Phi)=\frac{\gamma T}{4\pi\mu D(\Phi)
U(\Phi)}\,e^{-U(\Phi)/T}\,,
\label{3}
\ee
where
\be
U(\Phi)=T L^2(\Phi,\Phi)=\frac{1}{2}
\left[f_s(\Phi)-f_a(\Phi)\right]^2~,
\label{4}
\ee
and
$D(\Phi)= (1/\sqrt{2\pi T_d}) d\,
f_s(\Phi)/d\Phi=
\exp\{\text{irf\,}^2(2\Phi-1)\}$.
Notice that $U(\Phi)$ in (\ref{4}) has a clear physical
interpretation. It is the energy
barrier between the actual force $f_a(\Phi)$ (\ref{fa}) and the front
value $f_s(\Phi)$ (\ref{gjkwl}) of the distribution $Q(f,t)$ in (\ref{fgjaa}).
Equation (\ref{3}) is valid as long as
$U(\Phi) \gg T$, which implies $\Phi < \Phi_c$, where
$\Phi_c$ is such that the force $f_a$ per surviving fiber reaches the
average strength $1$: $f_a(\Phi_c)=1$ yielding $\Phi_c= 1-f_0$.
Such fractions $\Phi \to \Phi_c$ correspond to the ultimate regime
of explosive failure. We have checked by direct
numerical calculations that (\ref{3})
provides an exceedingly precise approximation of equation
(\ref{mgler})
as long as $\mu \gtrsim 1$ and not too close to $\Phi_c$ (in practice
to within a few percent).

To go further, we need to distinguish between two regimes,
$\Phi < \Phi^*$ and $\Phi > \Phi^*$, where $\Phi^*$, solution of
$dR(\Phi)/d\Phi=0$, corresponds to the minimum failure rate.
For $\mu \gtrsim 1$,
$\Phi^*$ is actually independent of the temperature $T$ and is the root of 
the
equation
\be
D(\Phi^*) (1-\Phi^*)^{2} = \alpha~, ~~~~~
\alpha=\frac{f_0}{\sqrt{2\pi T_d}}~,
\label{19}
\ee
where $\alpha$ is a important physical parameter
quantifying the strength of the disorder relative to the applied force.
The explicit approximate solution of equation (\ref{19}) is
\be
\Phi^*(\alpha)=\begin{cases}
\frac{20-\pi-4(4-\pi)\alpha}
{24-2\pi+8(4+\pi)\alpha}\,, & \alpha<3/2\,,\\[1mm]
\frac{1}{2}\,
\text{erfc\,}\left(\sqrt{\ln\alpha}\right) \frac{\alpha
\sqrt{\pi\ln\alpha}}{1+\alpha\sqrt{\pi\ln\alpha}}\,, &
\alpha>3/2\,.
\end{cases}
\label{approx}
\ee
For example, this gives $\Phi^*(\alpha=2)=0.089$ compared
with the exact value $0.092$.

It follows from (\ref{3}) that the time to reach some
$\Phi$ is given by
$\gamma\,T\,t=4\pi\mu\int_0^\Phi
D(z)U(z)\, e^{U(z)/T}dz$.
For $0<\Phi<\Phi^*(\alpha)$,
due to the exponential factor $e^{U/T}$, the main
contribution to the last integral comes from a small
neighborhood of the upper integration limit. This yields
\be
\gamma\,t\simeq 4\pi\mu
\frac{D(\Phi)U(\Phi)}
{A(\Phi)}\,e^{U(\Phi)/T}\,, ~~
A(\Phi)=\frac{dU(\Phi)}{d\Phi}~,
\label{15}
\ee
for $\Phi<\Phi^*$.
This approximation is correct under the assumption that
$e^{U(\Phi)/T}$ is rapidly increasing with $\Phi$, i.e., if
$\left|A(\Phi)\right|\gg T$.
The absolute value $|\dots|$ stresses that this condition
applies also for $\Phi > \Phi^*(\alpha)$. The same
reasoning in this case gives a similar approximation
\be
\gamma (t_c-t) \simeq 4\pi\mu
\frac{D(\Phi)U(\Phi)}
{\left|A(\Phi)\right|}\,
e^{U(\Phi)/T}\,, \qquad \Phi>\Phi^*\,,
\label{18}
\ee
where $t_c-t$ is the time to complete rupture. The
condition $\left|A(\Phi)\right|\gg T$ shows that both relations
(\ref{15}) and (\ref{18}) do not work in the vicinity of the
minimum rate of fiber failures given by the solution of (\ref{19}), for
which $A(\Phi^*)=0$.
For $0<\Phi<\Phi^*$, combining (\ref{3}) and (\ref{15}), we obtain
$\dot{U}=T/t$ for $t<t^*$,
where $t^*$ is defined by $\Phi(t^*)=\Phi^*$.
This gives
\be
U\left[\Phi(t)\right]=T\ln\gamma t \qquad
(t<t^*)\,
\label{2.3}
\ee
valid for $\gamma t \gg 1$ (this condition simply means that the
thermal fluctuations have had time to contribute
several independent jolts). The constant of integration
gives the $\ln \gamma$ contribution determined from
matching with the initial stage.
Replacing the lhs of (\ref{4})
by (\ref{2.3}) and putting $f_a \simeq f_0$ (for $\Phi$ small)
gives, in view of equation (\ref{3}), the fraction rate
\be
\dot{\Phi}\simeq\frac{1}{4\pi\mu\,t\,\ln\gamma t}\,
\exp\left[-\frac{1}{2 T_d}\left(1-f_0- \sqrt{2T\ln\gamma
t}\right)^2\right]\,.
\label{2.7}
\ee
Expression (\ref{2.7}) is one of our main results: the failure rate
$\dot{\Phi}$ of
fibers decreases after application of the load
proportionally to $1/t$, up to logarithm corrections. This $1/t$ decay
lasts as long as $\Phi$ remains smaller than $\Phi^*$. This $1/t$ law is 
known
in seismology as the Omori law. It is also ubiquitous in creep experiments
with exponents that are often close to or smaller than our prediction $1$.
For intermediate times
such that $\gamma t < e^{1/2T}$, $\dot{\Phi}
\sim {1 \over t~\ln \gamma t}~e^{(1-f_0)\sqrt{2T\ln\gamma t}/T_d}$, which gives
an apparent exponent $\sim 1/t^p$ with $p<1$. For $\ln \gamma t \gg 
(1-f_0)^2/2T$, $p \to 1+(T/T_d)$ which is close to but slightly larger than $1$.
Numerical simulations
confirm these predictions accurately. See for instance figure 2 of
\cite{Politi}
which our theory explains quantitatively.
Exact numerical integration and our analytical approximation
coincide everywhere, excluding a time interval
corresponding to a very small vicinity of the stationary point $\Phi^*$.
Note that Andrade's law \cite{Andrade} also derives from the deformation 
rate
being proportional to $df_a(t)/dt = f_0
\dot{\Phi}/(1-\Phi(t))^2\propto \dot{\Phi}$
as $\Phi(t)$ varies much more slowly than $\dot{\Phi}$.

Let us now turn to the description of the second regime
$\Phi(t)>\Phi^*$, relevant to obtain the failure rate up to global failure.
Combining (\ref{3}) and (\ref{18}), we obtain the expression
\be
U\left[\Phi(t)\right]=T\ln [\gamma(t_c-t)]~~~~~
(\Phi^*<\Phi < \Phi_c)~,
\label{second}
\ee
which is analogous to (\ref{2.3}).
The regime $\Phi^*<\Phi(t)<\Phi_c$ is strongly influenced by
thermal fluctuations, so that the disorder term can be neglected to obtain,
in view of (\ref{second}) and (\ref{4}),
$\Phi_c-\Phi(t)=f_0\,\frac{\sqrt{2 T\ln\gamma (t_c-t)}}{1-\sqrt{2
T\ln\gamma (t_c-t)}}$.
Differentiating both sides of this expression with respect to
$t$ yields the failure rate
\be
\dot{\Phi}(t)=C(t)/(t_c-t)\,,
\label{3.6}
\ee
where $C(t) = f_0\,T/[c(1-c)^2]$ with
$c=\sqrt{2 T\ln [\gamma (t_c-t)]}$.
This is the second important result of our analysis
(see also eq.(B11) in \cite{cilietal1}), which shows that,
for $\Phi>\Phi^*$, the failure rate accelerates towards a finite-time
singularity approximately as $\sim 1/(t_c-t)$. Such a behavior has been
documented extensively in experiments on rupture of heterogeneous material
\cite{samsor}. Our analysis provides a novel mechanism for the ubiquitous
time-to-failure regime observed in heterogeneous material. Strong
quenched heterogeneity
has been shown to play an essential role in controlling the critical nature
of the rupture process \cite{Leungetal} and in the existence of a
time-to-failure power law such as (\ref{3.6}). Here, we confirm that the
heterogeneity is essential to renormalize the thermal fluctuations
\cite{cilietal1,Politi}. While
the philosophy is similar, the mechanism is different and novel.
As for the Omori law, the logarithmic corrections in (\ref{3.6})
may give an apparent exponent of the power law, slightly smaller than $1$,
as observed in experiments. Our numerical tests show that expression
(\ref{3.6})
provides an approximation which coincide almost everywhere with the exact
solution inside the interval $t^* < t < t_c$.

There is a simple physical interpretation of the transition between
the two above mentioned rate behaviors (\ref{2.7}) and (\ref{3.6}).
To explain the first (rate decaying) regime, consider
the degenerate case of spontaneous fracture ($f_0=0$,
$\Phi^*=1$) for which
$U(\Phi)=f_s^2/2$. As time increases, $f_s$ grows, the
remaining fibers are stronger and the failure rate decays
together with the rate of change of the energy barrier.
The second regime can be qualitatively understood by
taking the limit of zero disorder ($T_d=0$,
$\Phi^*=0$), leading to $U(\Phi)=(1-f_a)^2/2$.
The force $f_a$ per remaining fiber grows
with time, the fibers break more and more easily and the failure rate grows
to give the fracture in finite-time.
In the intermediate case $0<\Phi^*<\Phi_c$,  due to the
competition between the growth
of $f_s$ and $f_a$, the two regimes co-exist. At early
times, the growth of $f_s$ dominates giving
the Omori and Andrade laws, followed by the
growth of $f_a$ in the second regime $\Phi>\Phi^*$ giving
the power law finite-time singularity.

Lastly, we turn to the behavior for $\Phi \approx \Phi^*$, which turns
out to provide the dominant contribution for the total time for rupture,
as shown in \cite{cilietal1,Politi}. Indeed, the fiber bundle spends most of
its time in the vicinity of the stationary point $\Phi^*$, corresponding to
the minimum failure rate.
In this case, $U$ can be expanded as
\begin{equation*}
U(\Phi)=U(\Phi^*)-B(\Phi^*)(\Phi-\Phi^*)^2,~~
B(\Phi)=-\frac{1}{2}\frac{d^2 U(\Phi)}{d\Phi^2},
\end{equation*}
and equation (\ref{3}) becomes
\be
\dot{\Phi}\simeq R(\Phi^*)\exp\left[-\frac{B(\Phi^*)}{T}
\left(\Phi-\Phi^*\right)^2\right]\,.
\label{4.3}
\ee
The solution of this equation is
\be
\text{erfi\,}\left(\sqrt{\frac{B(\Phi^*)}{T}}\,(\Phi-\Phi^*)\right)=
2\, \sqrt{\frac{B(\Phi^*)}{\pi T}}\,
R(\Phi^*)\,(t-t^*)\,,
\label{4.4}
\ee
where $\text{erfi\,}(z)=\frac{1}{i}\,\text{erf\,}(iz)$
is the imaginary error function. Using its asymptotics
$\text{erfi\,}(z)\sim \frac{1}{\sqrt{\pi} z}\,e^{z^2}$ for
large $z$ together with (\ref{4.4}), equation (\ref{4.3}) becomes
\begin{equation*}
\frac{d\Psi^2}{dt}\simeq
\frac{T}{B(\Phi^*)\, (t-t^*)}~,~~~\Psi=\Phi-\Phi^*\,,
\end{equation*}
whose solution yields the fracture rate
\begin{equation*}
\dot{\Phi}\simeq \frac{\sqrt{T}\text{sign\,}(t-t^*)}{2|t-t^*|
\sqrt{B(\Phi^*)\,\ln(\gamma|t-t^*|)}}~~~~~
(\gamma|t-t^*|\gg 1)~.
\end{equation*}
Expression (\ref{4.3}) allows us additionally to calculate
the total lifetime $t_c$ of the fiber bundle:
\be
\gamma\,t_c=\frac{1}{R\left(\Phi^*\right)}
\int_{-\infty}^\infty \exp\left[-
\frac{B(\Phi^*)}{T}
\left(\Phi-\Phi^*\right)^2\right]\,d\Phi\,.
\ee
The calculation of this integral with the use of (\ref{3}) gives
\be
\gamma\,t_c\simeq 4\pi\sqrt{\pi}\,
\frac{D(\Phi^*)
U(\Phi^*)}{\sqrt{B(\Phi^*)}}\, \frac{\sqrt{T_d}}{T}\,
\exp\left[-\frac{U(\Phi^*)}{T}\right]\,.
\label{4.8}
\ee
This expression, together with (\ref{approx}),
recovers the main result of \cite{cilietal1,Politi},
while improving on the prefactors to the main Arrhenius-type dependence.

\begin{figure}[h]
\includegraphics[width=8cm]{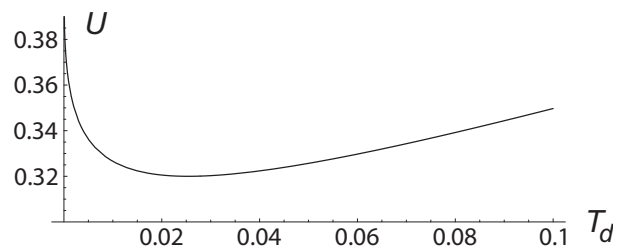}
\caption{\label{Fig1} Effective barrier energy
$U(\Phi^*)$ as a function of the disorder strength $T_d$, in the case
where $\Phi^*=1/2$, corresponding to $T_d^*=\frac{8}{\pi}\,f_0^2$,
for $f_0=0.1$ leading to $T_d^*\simeq 0.025$.
This illustrates the non-monotonous behavior of
$U(\Phi^*)$ and thus of the lifetime $t_c$ with $T_d$.}
\end{figure}

Using (\ref{4}) and (\ref{19}), $U(\Phi^*)$ can be written explicitly
\be
U(\Phi^*)=\left[
\frac{\Phi_c-\Phi^*}{1-\Phi^*}
\pm
\sqrt{\ln\left(\frac{\alpha}{(1-\Phi^*)^2}\right)}
\right]^2
\label{4.13}
\ee
where the sign $+$ (resp. $-$) corresponds to the case $\Phi^*>1/2$ 
(resp. $\Phi^*<1/2$). As shown in Fig.~\ref{Fig1}, 
$U(\Phi^*)$ is a non-monotone function of $T_d$.
Due to the above mentioned competition between quenched disorder
and the growth of the actual force $f_a$, $U(\Phi^*)$ decreases as long as
$T_d<T_d^*$ and then increases with
increasing $T_d$ beyond $T_d^*$. The first regime $T_d<T_d^*$
corresponds to the effect discovered in Refs.~\cite{cilietal1,Politi}
and mentioned above
of the renormalization of thermal fluctuations by quenched disorder, and
consequently of decreasing strength by increasing the disorder.
Since a larger $U(\Phi^*)$ corresponds to a large lifetime through (\ref{4.8}),
we uncover the new effect of a strengthening of the fiber system
by increasing the disorder beyond a certain threshold.
All our formulas have been checked by direct numerical integration with
excellent agreements. We expect that extensions of the DFBM to
non-mean field power law interactions \cite{Kunetal} will not change
our results
qualitatively but may modify the Omori's and time-to-failure exponents.

We acknowledge useful exchanges with S. Ciliberto and A. Politi.


\vskip -0.5cm
\begin{thebibliography}{}

\bibitem{Andrade} E.N. da C. Andrade, Proc. R. Soc. London A 84, 1
(1910); 90, 329 (1914).

\bibitem{Omori} T. Utsu, Y. Ogata and S. Matsu'ura,
J. Phys. Earth 43, 1-33 (1995).

\bibitem{cilietal1} S. Ciliberto, A. Guarino and R. Scorretti,
Physica D 158, 83-104 (2001).

\bibitem{Politi} A. Politi, S. Ciliberto and R. Scorretti,
Phys. Rev. E 66, 026107 (2002).

\bibitem{early} R.I.B. Selinger, Z. Wang and W.M. Gelbart,
Phys. Rev. A 43, 4396 (1991) and references therein;
A. Buchel and J.P. Sethna, Phys. Rev. Lett. 77, 1520 (1996).

\bibitem{samsor} J.-C. Anifrani et al.,
J.Phys.I France 5, 631-638 (1995); A. Garcimartin et al., Phys. Rev.
Lett. 79, 3202 (1997),
see also S.G. Sammis and D. Sornette, Proc. Nat. Acad. Sci. USA 99
SUPP1, 2501-2508 (2002)
for a review and references therein.

\bibitem{Leungetal} J.V. Andersen. D. Sornette and K.-T. Leung,
Phys. Rev. Lett. 78, 2140-2143 (1997).

\bibitem{Kunetal} F. Kun, Y. Moreno, R.C. Hidalgo and H.J. Herrmann,
Europhys. Lett. 63, 347-353 (2003).


\end{thebibliography}
\end{document}